\documentclass[11pt]{article}
\usepackage{graphicx}
\usepackage{amsmath}
\usepackage{amsfonts}
\usepackage{epsfig}

\usepackage{epstopdf}
\usepackage{color}

\newcommand{\be}{\begin{equation}}
\newcommand{\ee}{\end{equation}}
\newcommand{\bes}{\begin{equation*}}
\newcommand{\ees}{\end{equation*}}
\newcommand{\beqn}{\begin{eqnarray}}
\newcommand{\eeqn}{\end{eqnarray}}
\newcommand{\beqns}{\begin{eqnarray*}}
\newcommand{\eeqns}{\end{eqnarray*}}

\newcommand{\lkr}{\left(}
\newcommand{\rkr}{\right)}
\newcommand{\lkv}{\left[}
\newcommand{\rkv}{\right]}
\newcommand{\lfi}{\left\{}
\newcommand{\rfi}{\right\}}
 
\newcommand{\EE}{\ensuremath{{\mathbb E}}}
\newcommand{\II}{\ensuremath{{\mathbb I}}}

\newcommand{\NN}{\ensuremath{{\mathbb N}}}

\newtheorem{lemma}{Lemma}
\newtheorem{theorem}{Theorem}

\newcommand{\fr}[1]{(\ref{#1})}

\newcommand{\vart}{\vartheta}

\newcommand{\ph}{\varphi}
\newcommand{\del}{\delta}

\newcommand{\eps}{\varepsilon}
\newcommand{\Ga}{\Gamma}
\newcommand{\ga}{\gamma}
\newcommand{\te}{\theta}
\newcommand{\om}{\omega}
\newcommand{\lam}{\lambda}

\newcommand{\up}{\upsilon}

\newcommand{\sig}{\sigma}

\newcommand{\Om}{\Omega}
\newcommand{\Te}{\Theta}
\newcommand{\Del}{\Delta}

\newcommand{\bx}{\mathbf{x}}
\newcommand{\bg}{\mathbf{g}}
\newcommand{\bq}{\mathbf{q}}
\newcommand{\bu}{\mathbf{u}}

\newcommand{\bh}{\mathbf{h}}

\newcommand{\bA}{\mbox{$\mathbf A$}}

\newcommand{\bG}{\mathbf{G}}

\newcommand{\bom}{\mbox{\mathversion{bold}$\om$}}
\newcommand{\bte}{\mbox{\mathversion{bold}$\te$}}

\newcommand{\Tr}{\mbox{Tr}}
\newcommand{\card}{\mbox{card\,}}
\newcommand{\Var}{\mbox{Var}}
\newcommand{\AIF}{\mbox{AIF}}
\newcommand{\vect}{\mbox{vec}} 
\newcommand{\wind}{\mbox{wind}}
\newcommand{\SNR}{\mbox{SNR}}

\newcommand{\Cov}{\mbox{Cov}}
\newcommand{\std}{\mbox{std}\,}

\newcommand{\etal}{{\it et al.\ }}

\long\def\ignore#1{}

\title{\Large{\bf Anisotropic functional Laplace deconvolution }}

\author{
\large{ Rida Benhaddou }   \\  
Department of Mathematics, Ohio University\\ \\  \large{Marianna Pensky and Rasika Rajapakshage} \\ 
  Department of Mathematics, University of Central Florida } 

\date{}

\begin{document}

\maketitle
\begin{abstract}
In the present paper we consider the problem of estimating a three-dimensional function $f$ based on observations 
from its noisy Laplace convolution. Our  study is motivated by  the analysis of Dynamic Contrast Enhanced (DCE) imaging data. 
We construct an adaptive wavelet-Laguerre estimator of $f$, derive minimax lower 
bounds for the $L^2$-risk when $f$ belongs to a three-dimensional Laguerre-Sobolev ball and demonstrate that 
the wavelet-Laguerre estimator is adaptive and asymptotically near-optimal  
in a wide range of Laguerre-Sobolev spaces. We carry out a limited simulations study and show that the estimator
performs well in a finite sample setting. Finally, we use the technique for the solution 
of the Laplace deconvolution problem on the basis  of DCE Computerized Tomography data.
\vspace{2mm} 

{\bf  Keywords and phrases}: {functional Laplace deconvolution, minimax convergence rate, Dynamic Contrast Enhanced  imaging   }

\vspace{2mm}

{\bf AMS (2000) Subject Classification}: {Primary: 62G05,  Secondary: 62G08,62P35  }
\end{abstract}


\section {Introduction }
\label{sec:introduction}
\setcounter{equation}{0} 

Consider  an equation
\be \label{eq1}
Y(t, \bx)= q(t, \bx) + \eps \xi(t, \bx) \quad \mbox{with} \quad 
q(t, \bx)= \int^{t}_{0} g(t-z) f(z, \bx) dz.
\ee
where $\bx = (x_1, x_2)$, $(t, x_1, x_2) \in U=[0,\infty) \times [0,1]\times [0,1]$ and $\xi(z, x_1, x_2)$ 
is the three-dimensional Gaussian white noise
such that 
$$
\Cov \lfi \xi(z_1, x_{11}, x_{12}), \xi(z_2, x_{21}, x_{22}) \rfi = 
\II(z_1 = z_2) \, \II(x_{11} = x_{2 1})\, \II(x_{12} = x_{22}).
$$
Here and in what follows, $\II(A)$ denotes the indicator function of a set $A$.
Formula \fr{eq1} can be viewed as  a noisy version of a functional Laplace  convolution equation.
Indeed, if   $\bx$ is fixed, then \fr{eq1} reduces to a noisy version of the Laplace  convolution equation 
\be \label{eq2}
Y(t)= q(t) + \eps \xi(t) \quad \mbox{with} \quad  q(t) =  \int^{t}_{0} g(t-z) f(z) dz,
\ee
 that was  recently studied by Abramovich \etal (2013), Comte \etal (2017) and Vareschi (2015).


Equation \eqref{eq1} represents a  white-noise version of the Laplace convolution equation
which corresponds to the  observational version of the equation 
\be \label{eq:observ_mod}
Y(t_i, x_{1,j},x_{2,l})=  \int^{t_i}_{0} g(t_i -z) f(z, x_{1,j},x_{2,l}) dz + \sig \xi_{i,j,l},   
\ee
where $i=1, \cdots,n_0,$ $j=1, \cdots, n_1,$ $l=1, \cdots, n_2,$
$t_i = i T/n_0$ are equispaced on the interval $[0,T]$, $x_{1,j} = j/n_1$ and $x_{2,l} = l/n_2$ and 
$\xi_{i,j,l}$ are standard normal variables that are independent for different $i,j$ and $l$. If 
$n_0, n_1$ and $n_2$ are large, then equation \fr{eq1} serves as an ``idealized'' version of equation 
\fr{eq:observ_mod}.  This result is rigorously proved in the  case of the Gaussian regression model 
(see,  e.g. Brown and Low (1996)), and  it is well known that it holds for a large variety of settings. 
Abramovich \etal (2013) studied a one-dimensional $(n_1=n_2=1)$  version of the equation \fr{eq:observ_mod}.   
It follows from the upper and lower bounds in their paper that the correspondence between equations 
\fr{eq2} and the one-dimensional version of equation \fr{eq:observ_mod} holds with 
$\eps = \sig T/\sqrt{n}$ where $n = n_0 n_1 n_2$ (since $n_1=n_2=1$).

Comte \etal (2017) also studied solution of equation \fr{eq:observ_mod} in the case of $n_1=n_2=1$
and rigorously investigated the implications of the fact that observations are taken on the finite interval $[0,T]$
rather  than  on the positive part of the real line. They showed that the latter leads to a much more involved mathematical arguments. 
On the other hand, Vareschi (2015) considered   equation \fr{eq2} and, building upon an 
earlier version of Comte \etal (2017), derived the lower and the upper bounds for the error in the white noise version 
of the Laplace deconvolution problem.  Our paper can be regarded as an extension of Vareschi's (2015)
results to the case when Laplace convolution equation has a spatial component and the function of interest is anisotropic, i.e.,
may have different degrees of smoothness in different directions. Therefore, our objective is to show how 
utilizing the spatial smoothness of the unknown function $f$ leads to its more precise recovery.

 
 Our study is motivated by  the analysis of Dynamic Contrast Enhanced (DCE) imaging data. 
DCE imaging provides a non-invasive measure of tumor angiogenesis and has great potential 
for cancer detection and characterization, as well as for monitoring, \textit{in vivo}, 
the effects of therapeutic treatments (see, e.g., Bisdas {\it et al.}  (2007), 
Cao (2011); Cao {\it et al.}  (2010) and   Cuenod {\it et al.}  (2011)).  
The common feature of DCE imaging techniques is that each of them uses the rapid injection 
of a single dose of a bolus of a contrast agent  and monitors its progression in the vascular network 
by sequential  imaging at times $t_i$, $i=1, \cdots, n$. This is accomplished  by measuring the 
pixels' grey levels that are proportional to the   concentration of the contrast agent in the corresponding voxels.
At each time instant $t_i$, one obtains an image of an artery as well as  a collection 
$Y (t_i, \bx)$ of measurements for each voxel $\bx$. For example, in the case of a CT scan, $Y (t_i, \bx)$
are the  Hu units  which represent  the opacity of the material to X-rays.  The images of the artery 
allow  to estimate the  so called  Arterial Input Function, $\AIF(t)$,  which quantifies the total amount of the contrast agent
entering the area of interest.
Comte \etal (2017) described the DCE imaging experiment in great detail and showed that
the cumulative distribution function $F(z, \bx)$ of the sojourn times for the particles of the contrast agent
entering a tissue voxel $\bx$ satisfies the following equation
\be \label{complete_DCE}
 Y (t, \bx) = \int_0^{t-\delta} g (t-z)\, \beta(\bx) (1 - F(z, \bx)) dz + \eps \xi(t, \bx).
\ee
Here the errors $\xi(t, \bx)$ are independent for different $t$ and $\bx = (x_1, x_2)$,  
$g(t) =  \AIF(t)$,    a positive coefficient $\beta (\bx)$ is related to a fraction of the contrast agent
entering the   voxel $\bx$ and  $\del$ is the time delay that can be easily estimated from data. 
The function of interest is $f(z, \bx) = \beta(\bx) (1 - F(z, \bx))$ where the distribution function 
$F(z, \bx)$ characterizes the properties of the tissue voxel $\bx$ and can be used 
as the foundation for medical conclusions.

Since the Arterial Input Function can be estimated by denoising and averaging the observations over all 
voxels of the aorta, its estimators incur much lower errors than those of  the left hand side of equation 
\fr{complete_DCE}. For this reason, in our theoretical investigations, we shall treat function $g$ in  \fr{complete_DCE} as known.
In this case, equation  \fr{complete_DCE} reduces to the form \fr{eq1} that we study  in the present paper.
If one is interested in taking the uncertainty about $g$ into account, this can be accomplished using 
methodology of Vareschi (2015).

 Laplace deconvolution equation \fr{eq2}  was first studied in Dey \etal(1998) under the assumption that 
$f$ has $s$  continuous derivatives on $(0, \infty)$. However, the authors only considered a very specific kernel, $g(t)=be^{-at}$,
and assumed that $s$ is known, so their estimator  was not adaptive.
 Abramovich \etal (2013) investigated Laplace deconvolution based on discrete noisy data. 
They implemented the kernel method with the bandwidth selection carried out by the Lepskii's method. 
The shortcoming of the approach is that it is strongly dependent on the exact knowledge of the kernel $g$.  
Recently,  Comte \etal~(2017)  suggested a method which  is based on the expansions of the kernel, 
the unknown function $f$ and the observed signals over Laguerre functions 
basis. This  expansion results  in an infinite system of linear equations with the   lower triangular  Toeplitz matrix. 
The system is then truncated and the  number of terms  that are kept  in the series expansion of the estimator 
is controlled via a complexity penalty. One of the advantages of the technique is that it considers a more realistic  setting
where $Y(t)$  in equation \fr{eq2} is observed at discrete time instants on an interval $[0,T]$ 
with $T < \infty$ rather than at every value of $t$. Finally, Vareschi (2015) derived a minimax optimal 
estimator of $f$ by thresholding the Laguerre coefficients in the expansions when $g$ is unknown and is measured with noise.

In the present paper, we consider the functional version \fr{eq1} of the Laplace convolution equation \fr{eq2}.
The study is motivated by the DCE imaging problem  \fr{complete_DCE}. Due to the high level of noise 
in the left hand side of  \fr{complete_DCE}, a voxel-per-voxel recovery of individual curves is highly inaccurate.
For this reason, the common approach is to cluster the curves for each voxel and then to average the curves 
in the clusters (see, e.g., Rozenholc and Rei\ss{}  (2012)). As the result, one does not recover individual 
curves but only their cluster averages. In addition,  since it is impossible to assess the clustering errors, 
the estimators may be unreliable even when estimation errors are small.  On the other hand, the functional approaches,
in particular, the wavelet-based techniques, allow to denoise a multivariate function of  
interest while still preserving its significant features.

The objective of this paper is to solve the functional Laplace deconvolution problem \fr{eq1} directly. 
In the case of the Fourier deconvolution problem,   Benhaddou \etal (2013) 
demonstrated that the functional deconvolution solution usually has a much better precision compared  
to a combination of   solutions of separate convolution equations. Below we adopt some of the ideas 
of Benhaddou \etal (2013)  and apply them to the solution of the functional Laplace convolution equation.
Specifically,  we assume that the unknown function belongs to an anisotropic   Laguerre-Sobolev space and 
recover it using a combination of wavelet and Laguerre functions expansion. Similar to Comte \etal (2017),
we expand the kernel $g$ over the Laguerre basis  and $f(t, \bx)$, $q(t, \bx)$ 
and $Y(t, \bx)$ over the Laguerre-wavelet basis and carry out denoising 
by thresholding the coefficients of the expansions, which naturally leads to truncation of the infinite system of equations 
that results from the process. We derive the minimax lower bounds for the $L^2$-risk in the model \fr{eq1}  
and  demonstrate that the wavelet-Laguerre estimator is adaptive and asymptotically near-optimal within a 
logarithmic factor in a wide range of Laguerre-Sobolev balls. We carry out a limited simulation study and 
then finally apply our technique to recovering of $f(z, \bx)$ in equation \fr{complete_DCE} on the bases 
of DCE-CT data.

Although, for simplicity, we only consider the white noise model for the functional Laplace convolution equation 
\fr{eq1}, the theoretical results can be easily generalized to its observational 
version \fr{eq:observ_mod} by following Comte \etal (2017). However, as it is evident from Comte \etal (2017),
the latter will lead to much more complex calculations and will make the paper very difficult to read while adding 
very little to the paper conceptually. For this reason, in the present paper, we avoid this extension.

The rest of the paper is organized as follows. In Section~2, we describe the construction of 
the wavelet-Laguerre estimator for $f(t, \bx)$ in equation \fr{eq1}. In Section~3, we derive 
the minimax lower bounds for the $L^2$-risk for any estimator of $f$  in \fr{eq1} over anisotropic   
Laguerre-Sobolev balls. In Section~4, we demonstrate that the wavelet-Laguerre estimator 
is adaptive and asymptotically minimax near-optimal (within a logarithmic factor of $\eps$)   
in a wide range of Laguerre-Sobolev balls. Section~5  presents a limited simulation study followed by a real data example
in Section~6. The proofs of the statements of the paper are placed in the Section~7. 
Finally, Section~8 provides some supplementary results from the theory of banded Toeplitz matrices.


\section {Estimation Algorithm. }
\label{sec:est-al}
\setcounter{equation}{0}

In what follows we are going to use the following notations.
Given a matrix $\mathbf A$, let  $\mathbf A^T$ be the transpose of $\mathbf A$,
$\|\bA\|_F = \sqrt{\Tr(\bA^T \bA)}$ and $\|\bA\| = \lambda_{\max} (\bA^T \bA)$ be, respectively,
the Frobenius and the spectral norm  of a matrix $\bA$, where $\lambda_{\max} (\mathbf U)$ is the largest,
 in absolute value, eigenvalue of $\mathbf U$. We denote by $[\mathbf A]_m$ the upper left $m\times m$
sub-matrix of $\mathbf A$. Given a vector $\bu \in {\mathbb R}^k$, we denote by $\| \bu \|$ its Euclidean
norm and, for $p\leq k$, the $p\times 1$ vector with the first $p$ coordinates of $\bu$, by $[\bu]_p$. For any
 function $t \in L_2(\mathbb R_+)$, we denote by $\| t \|_2$ its $L_2$ norm on $\mathbb R_+$.
For vectors, whenever it is necessary, we use the superscripts to indicate  dimensions of the vectors
and subscripts to denote their components.  
Also, $a\vee b = \max(a,b)$ and $a \wedge b = \min(a,b)$.\\

Consider a finitely supported periodized $r_0$-regular wavelet basis (e.g., Daubechies)
$\psi_{j,k}(x)$  on $[0,1]$. Form a product wavelet basis $\Psi_{\bom} (\bx) = \psi_{j_1,k_1}(x_1) \psi_{j_2,k_2}(x_2)$
on  $[0,1] \times [0,1]$ where $\bom \in \Omega$ with 
\be \label{eq:setOm}
\Om = \lfi \bom= (j_1,k_1; j_2,k_2): j_1, j_2 = 0, \cdots, \infty;\ k_1 = 0, \cdots, 2^{j_1-1},\ k_2 = 0, \cdots, 2^{j_2-1} \rfi.
\ee 
Denote functional wavelet coefficients of $f(t,\bx)$, $q(t, \bx)$, $Y(t, \bx)$
and $\xi(t, \bx)$ by, respectively, $f_{\bom} (t)$, $q_{\bom} (t)$, $Y_{\bom} (t)$
and $\xi_{\bom} (t)$. Then, for any $t\in [ 0, \infty )$,  equation \fr{eq1} yields 
\be \label{eq:qtild}
Y_{\bom}(t)=  q_{\bom}(t) + \eps  {\xi}_{\bom}(t)\quad \mbox{with} \quad
q_{\bom}(t)= \int^t_0g(t-s) f_{\bom}(s)ds
\ee
and function $f(t, \bx)$ can be written as 
\be \label{ser}
f(t, \bx)= \sum_{\bom \in \Om}  {f}_{\bom}(t) \Psi_{\bom}(\bx), \quad 
f_{\bom}(t) = \int_{[0,1]^2}  f(t,\bx) \Psi_{\bom} (x) d\bx \quad \bx = (x_1, x_2).
\ee
Now, consider the orthonormal basis  that consists of a system of Laguerre functions 
\be \label{eq:Laguerre_func}
\ph_l(t) = e^{-t/2} L_{l} (t),\ \ l=0,1,2,  \ldots,
\ee
where $L_l(t)$ are Laguerre polynomials (see, e.g., Gradshtein and Ryzhik (1980), Section 8.97)
$$
L_l(t) = \sum_{j=0}^l (-1)^j {l \choose j} \frac{t^j}{j!},\ \ \ t \geq 0.
$$
It is known that functions  $\ph_l(\cdot)$, $l=0,1,2, \ldots$, form an orthonormal basis of the $L^2 (0, \infty)$ space
and, therefore, functions $ {f}_{\bom}(\cdot)$, $g(\cdot)$, $ {q}_{\bom}(\cdot)$    
and ${Y}_{\bom}(\cdot)$ can be expanded over this basis with coefficients
$ {\te}_{l; \bom}$, $g_l$, $ {q}_{l; \bom}$    
and $ {Y}_{l; \bom}$, $l = 1, \ldots, \infty$, respectively.
By plugging these expansions into formula \fr{eq:qtild}, we obtain the following
equation
\be \label{eq:qgf}
\sum^{\infty}_{l=0} q_{l; \bom}\, \ph_l(t) = \sum^{\infty}_{l=0} \sum^{\infty}_{k=0} 
   \theta_{l; \bom}\, g_k \  \int^t_0 \ph_k(t-s) \ph_l(s) ds.
\ee
Following Comte {\it et al.} (2017),  for each $\bom \in \Om$,  we represent 
coefficients of interest $\theta_{l; \bom}$, $l=0,1, \ldots,$  as a solution 
of an infinite triangular system of linear equations. Indeed, it is easy to check that 
(see, e.g., 7.411.4 in Gradshtein and Ryzhik (1980))
$$
\int_0^t \phi_k(x)  \phi_j(t-x)  dx =   e^{-t/2} \int_0^t L_k(x) L_j(t-x) dx =
\phi_{k+j}(t) - \phi_{k+j+1} (t).
$$
Hence,   equation \fr{eq:qgf} can be re-written as
$$
\sum_{k=0}^\infty q_{k; \bom}\, \ph_k(t) = \sum_{k=0}^\infty \, \lkv \theta_{k; \bom}\ g_0  +
\sum_{l=0}^{k-1}   (g_{k-l}  - g_{k-l-1})\, \te_{l; \bom} \rkv \, \ph_k(t).
$$
Equating coefficients for each basis function, we obtain an infinite triangular system of linear equations.
In order to use this system for estimating $f$, we choose a fairly large $M$ and define the following approximations of $f$ and $q$
based on the first $M$ Laguerre functions
\be \label{approx}
f_M(t, \bx)=  \sum_{\bom \in \Om} \sum_{l=0}^{M-1}  \theta_{l; \bom} \ph_l(t) \Psi_{\bom}(\bx), \quad 
q_M(t, \bx)=  \sum_{\bom \in \Om} \sum_{l=0}^{M-1} q_{l; \bom} \ph_l(t) \Psi_{\bom}(\bx).
\ee
%
Let $\bte^{(M)}_{\om}$, $\bg^{(M)}$ and $\bq^{(M)}_{\om}$ 
be $M$-dimensional vectors with elements $\te_{l; \bom}$, $g_l$ and $q_{l; \bom}$, $l=0,1, \ldots, M-1$,
respectively. Then, for any $M$ and any $\bom \in \Om$, one has $\bq^{(M)}_{\om} =  \bG^{(M)}  \bte^{(M)}_{\om}$ where
$\bG^{(M)}$ is the lower triangular Toeplitz matrix  with elements $G^{(M)}_{i,j}$, $0 \leq i,j \leq M-1$ 
\be  \label{Toeplitz_matr}
G^{(M)}_{i,j}  = \left\{
\begin{array}{ll}
 \, g_0, & \mbox{if}\ \   i=j,\\
 \,  (g_{i-j}  - g_{i-j-1}), & \mbox{if}\ \ j<i,\\
0, & \mbox{if}\ \ j>i.
\end{array} \right.
\ee
In order to recover  $f$ in \fr{eq1}, we estimate coefficients $q_{l; \bom}$ in \fr{approx} 
by 
\be \label{qhat}
\widehat{q}_{l; \bom}= \int^{\infty}_{0}   {Y}_{\bom}(t)\, \ph_l(t)\,dt, \quad l = 0,2, \ldots, 
\ee
and  obtain an estimator $\widehat{\bte^{(M)}_{\bom}}$ of vector $\bte^{(M)}_{\bom}$
of the form 
\be \label{hat_theta}
\widehat{\bte^{(M)}_{\bom}} =   (\bG^{(M)})^{-1}   \widehat{\bq^{(M)}_{\bom}}.
\ee
Denote by $\Om(J_1, J_2)$ a truncation of a set $\Om$ in \fr{eq:setOm}:
\be \label{eq:setOmJ}
\Om(J_1, J_2) = \lfi \bom= (j_1,k_1; j_2,k_2): 0 \leq j_i \leq J_i-1,\    k_i = 0, \cdots,  2^{j_i -1}; i=1,2  \rfi.
\ee
If we recovered $f$ from all its coefficients $\widehat{\bte^{(M)}_{\bom}}$
with $\bom \in \Om(J_1, J_2)$, the estimator would have a very high variance.
For this reason, we need to remove the coefficients that are not essential for representation of $f$.
This is accomplished by constructing a hard thresholding estimator for the function $f(t, \bx)$   
\be \label{fhat}
\widehat{f}(t, \bx)= 
\sum^{M-1}_{l=0}\  \sum_{\bom \in \Om(J_1, J_2)} \ \widehat{\theta}_{l; \bom} \, 
\II \left(| \widehat{\theta}_{l; \bom}|  > \lambda_{l, \varepsilon}  \right)\, \ph_l(t)\, \Psi_{\bom}(\bx),
\ee
where the values of $J_1$, $J_2$, $M$ and $\lambda_{l, \varepsilon}$ will be defined later.


\section {Minimax lower bounds for the risk. }
\label{sec:lower_bounds}
\setcounter{equation}{0}

In order to determine the values of parameters $J_1$, $J_2$, $M$ and $\lambda_{l, \varepsilon}$,
and to gauge the precision of the estimator $\widehat{f}$, 
we need to introduce some   assumptions on the function $g$. 
Let $r \geq 1$  be such that
\be \label{k_cond}
\left. \frac{d^j g(t)}{dt^j} \right|_{t=0} = \lfi
\begin{array}{ll}
0, & \mbox{if}\ \ j=0, ..., r-2,\\
B_r \ne 0, &  \mbox{if}\ \ j=r-1,
\end{array} \right.
\ee
with the obvious modification $g(0)=B_1 \ne 0$ for $r=1$.
We   assume that function $g(x)$ and 
its Laplace transform $G(s) =\int_0^{\infty} e^{-sx} g(x)dx$ satisfy the following conditions:
\\

 \noindent
{\bf Assumption A1.  }  $g \in L_1 [0, \infty)$ is $r$ times differentiable with   $g^{(r)} \in L_1 [0, \infty)$.
\\

 \noindent
{\bf Assumption A2. }  Laplace transform $G(s)$ of $g$ has no zeros with nonnegative 
real parts except for zeros of the form $s=\infty + ib$.
\\
 
\noindent
Assumptions 1 and 2 are difficult to check since their  verification relies on the exact knowledge of $g$
and the value of $r$. Therefore, in the present paper, we do not use the value of $r$ in our estimation algorithm
and aim at construction of an adaptive estimator that delivers the best   convergence rates that are  possible for 
the true unknown value of $r$ without its knowledge. Hence, we need to derive   the smallest  error 
that any estimator of $f$ can attain under Assumptions~A1 and A2.

For this purpose, we  consider  the generalized three-dimensional Laguerre-Sobolev ball of
radius $A$, characterized by its wavelet-Laguerre coefficients $\theta_{l; \bom} = \theta_{l; j_1, j_2, k_1, k_2}$ as follows: 
\be   \label{LagSob}
 {\cal B}^{s_1, s_2, s_3}_{\gamma, \beta}(A)=\left \{ f:  \sum^{\infty}_{l=0}\sum_{j_1=0}^\infty\sum_{j_2 = 0}^\infty 
2^{2js_1+2j's_2}  (l\vee 1)^{2s_3} \exp \lkr 2 \gamma\, l^\beta \rkr \,  \sum_{k_1=0}^{2^{j_1}-1}\, \sum_{k_2=0}^{2^{j_2}-1}\, 
\theta^2_{l; \bom} \leq A^2 \right \},
\ee 
where we assume that $\beta = 0$ if $\gamma = 0$ and    $\beta >0$ if  $\gamma >0$.

Note that if   $f$ were a function of $x$ and $y$ only, inequality \fr{LagSob} would contain only sums over $j_1$ and $j_2$ and 
would state that  function $f$ belongs to a two-dimensional Sobolev ball. On the other hand, the sum over $l$ provides upper bounds on the 
functional Laguerre coefficients. Observe that, unlike in the case of the wavelet coefficients   that are usually   bounded by powers of 
$2^{j_1}$ and $2^{j_2}$,  it is feasible for Laguerre coefficients to decrease exponentially with $l$ 
(see, e.g., Comte and  Genon-Catalot  (2015) for examples). Recall  also, that the original equation \fr{eq1} 
requires solution of an ill-posed problem in time variable (that corresponds to index $l$ in \fr{LagSob})
while represents functional regression in  space. The value of $r$ in Assumption {\bf A1} serves as the degree of ill-posedness 
 and, therefore,   affects only  the  precision of recovery of $f$ in the time but not the space domain.  
For this reason, in  the  expressions for the upper and the lower bounds of the error, the values of $s_1$ and $s_2$ 
are compared with $s_3/(2r)$ rather than $s_3$. 
In particular, in what  follows we shall assert that   both the lower and the upper bounds for the risk   
are expressed via  
\be \label{error_rates}
\Del(s_1,s_2,s_3,\ga,\beta,A) =  
\left\{ \begin{array}{ll} 
  A^2\ \left[A^{-2} \eps^2 \right]^{\frac{2s_1}{2s_1+1}} , & \mbox{if}\ \  s_1 \leq \min(s_2, s_3/(2r)),\ \gamma = \beta = 0\\
  A^2\   \left[A^{-2} \eps^2 \right]^{\frac{2s_2}{2s_2+1}} , & \mbox{if}\ \  s_2 \leq \min(s_1, s_3/(2r)),\ \gamma = \beta = 0\\
  A^2\ \left[A^{-2} \eps^2 \right]^{\frac{2s_3}{2s_3+2r}}, & \mbox{if}\ \ s_3 \leq \min(2r s_1, 2r s_2),\ \gamma = \beta = 0\\
  A^2\ \left[A^{-2} \eps^2 \right]^{\frac{2s_1}{2s_1+1}} , & \mbox{if}\ \  s_1 \leq s_2,\ \gamma >0, \beta > 0\\
  A^2\   \left[A^{-2} \eps^2 \right]^{\frac{2s_2}{2s_2+1}} , & \mbox{if}\ \  s_2 \leq s_1,\ \gamma >0, \beta > 0.\\
\end{array} \right.
\ee
In order to construct minimax lower bounds,   we define  the  maximum $L^2$-risk over the set $V$ of an estimator $\tilde{f}$ as
\be   \label{eq12}
 R_{\varepsilon}(\tilde{f}, V)= \sup _{f \in V}\, \EE \| \tilde{f} -f\|^2.
\ee 
The  following theorem provides the minimax lower bounds for the $L^2$-risk of any estimator $\tilde{f}$ of $f$.

\begin{theorem} \label{th:lowerbds}
Let $\min\{s_1, s_2  \} \geq 1/2$  and $s_3 \geq  1/2$ if $\gamma = \beta =0$. Then, if  $\eps$,
is small enough, under Assumptions A1 and A2,  for some absolute constant $\underline{C}>0$ independent of $\eps$, one has
 \be \label{lowerbds}
\inf_{\tilde{f}}\  R_{\varepsilon}(\tilde{f},\ {\cal B}^{s_1, s_2, s_3}_{\gamma, \beta} (A)) \geq \underline{C}\, \Del(s_1,s_2,s_3,\ga,\beta,A).
\ee
\end{theorem}

Note that the one-dimensional version \fr{eq2} of the problem \fr{eq1} corresponds to the situation when $s_1 = s_2= \infty$. 
Vareschi (2015) derived the upper and the lower bounds for the error in the case of $\gamma =0$. His lower bounds coincide 
with the lower bound given by \fr{error_rates} when $\ga=0$ and  $s_1 = s_2= \infty$.


\section {Upper bounds for the risk.  }
\label{sec:upper_bounds}
\setcounter{equation}{0}

In order to derive an upper bound for $R_{\varepsilon}(\widehat{f},\ {\cal B}^{s_1, s_2, s_3}_{\gamma, \beta} (A))$,
we need some auxiliary statements. 
Consider   $\bG^{(m)}$, the lower triangular Toeplitz matrix defined by formula \fr{Toeplitz_matr}
with $M=m$. The following results  follow  directly from Comte {\it et al.} (2017)
and Vareschi (2015).

\begin{lemma}\label{lem:Comte}
{\bf (Lemma 4, Comte   \etal  (2017), Lemma 5.4, Vareschi (2015)).}  
Let conditions A1 and A2 hold. Denote the elements of the last row of matrix $(\bG^{(m)})^{-1}$ by $\up_j$, $j=1, \cdots, m$.
Then, there exist  absolute positive constants $C_{G 1}$, $C_{G 2}$, $C_{\up 1}$ and  $C_{\up 2}$  independent of $m$
such that
\beqn \label{eq:norm_bounds}
C_{G 1}   m^{2r} & \leq & \, \| (\bG^{(m)})^{-1} \|^2    \leq  \| (\bG^{(m)})^{-1} \|_F^2  \leq   C_{G 2}    m^{2r},\\
\label{elem_bounds}
C_{\up 1}   m^{2r-1} & \leq & \,  \sum_{j=1}^m \up_j^2  \leq C_{\up 2}   m^{2r-1}.
\eeqn
\end{lemma}

\noindent
Using Lemma \ref{lem:Comte}, one  can obtain the following upper bounds for the errors 
of estimators $\widehat{\theta}_{l; \bom}$:

\begin{lemma} \label{lem:VarDev}
Let $\widehat{\theta}_{l; \bom}$ be the $l-th$ element of the vector  $\widehat{\bte^{(M)}_{\bom}}$ 
defined in \fr{hat_theta}. Then, under the Assumptions A1 and A2, one has 
\begin{align} \label{var}
& \Var\left[\widehat{\theta}_{l; \bom}\right]   \leq C_{\up 2}\  \varepsilon^{2}\, l^{2 r -1},\\
\label{mom4}
& \EE \left[\widehat{\theta}_{l; \bom} -  {\theta}_{l; \bom}\right]^4 \leq 3 C_{\up 2}^2\  \varepsilon^{4}\, l^{4 r -2},\\
\label{Largdev}
& \Pr \left(|\widehat{\theta}_{l; \bom} -  {\theta}_{l; \bom}| > \eps \sqrt{2 \nu 
\log(\eps^{-1}) \ l^{-1}}\ \| (\bG^{(l)})^{-1} \| \right) \leq \eps^\tau,
\end{align} 
provided $\nu \geq \tau C_{\up 2}/C_{G 1}$ where $C_{G 1}$  and $C_{\up 2}$   are defined in  
\fr{eq:norm_bounds} and \fr{elem_bounds}, respectively.
\end{lemma}

\noindent
Following Lemma \ref{lem:VarDev} we choose  $J_1$, $J_2$,  $M$   such that 
\be \label{Lev:J}
2^{J_1} = 2^{J_2} =A^2 \eps^{-2},\quad
M = \max\left\{ m \geq 1:\ \| (\bG^{(m)})^{-1} \| \leq \eps^{-2} \right\},
\ee
and thresholds $\lambda_{l, \varepsilon}$ of the forms
\be \label{Thres}
\lambda_{l, \varepsilon}   = 2\eps\ \sqrt{ 2\, \nu   \log(\eps^{-1}) \ l^{-1}}\ \| (\bG^{(l)})^{-1} \|,  
\ee
where the value of $\nu$ is large enough, so that it satisfies the inequality
\be    \label{nu-value}
\nu \geq  12 C_{\up 2}/C_{G 1},
\ee 
and $C_{\up 2}$ and $C_{G 1}$  and $C_{\up 2}$   are defined in  
\fr{eq:norm_bounds} and \fr{elem_bounds}, respectively.
Then, the following statement holds.

\begin{theorem} \label{th:upperbds}
Let $\min\{s_1, s_2  \} \geq 1/2$  and $s_3 \geq  1/2$ if $\gamma = \beta =0$. 
Let $\widehat{f}(t,\bx)$ be the wavelet-Laguerre estimator defined in \fr{fhat}, 
with $J_1$, $J_2$ and $M$ given by \fr{Lev:J}. Let  $A > 0$, and let condition \fr{LagSob} hold. 
If $\nu$ in \fr{Thres} satisfies inequality \fr {nu-value},
then,   under Assumptions A1 and A2, if  $\eps$,
is small enough, for some absolute constant $\overline{C}>0$ independent of $\eps$, one has
\be
R_{\widehat{f},\ \varepsilon}(B^{s_1, s_2, s_3}_{\gamma, \beta} (A)) \leq 
\overline{C}\, \Del(s_1,s_2,s_3,\ga,\beta,A)\, \left[\log(1/\eps) \right]^d,
\ee
where $\Del = \Del(s_1,s_2,s_3,\ga,\beta,A)$ is defined in \fr{error_rates} and 
\bes
d = \left\{ \begin{array}{ll}
 2s_1/(2s_1+1) + \II(s_1=s_2) + \II(s_3=2rs_1), & 
\mbox{if}\ \  s_1 \leq \min(s_2, \frac{s_3}{2r}),\ \gamma = \beta = 0\\
 2s_2/(2s_2+1) + \II(s_1=s_2) + \II(s_3=2rs_2), &
\mbox{if}\ \  s_2 \leq \min(s_1, \frac{s_3}{2r}),\ \gamma = \beta = 0\\
2s_3/(2s_3+2r) + \II(s_3=2rs_1) + \II(s_3=2rs_2), &
\mbox{if}\ \ s_3 \leq \min(2r s_1, 2r s_2),\ \gamma = \beta = 0\\
2s_1/(2s_1+1) + \II(s_1=s_2), &
\mbox{if}\ \  s_1 \leq s_2,\ \gamma >0, \beta > 0\\
 2s_2/(2s_2+1) + \II(s_1=s_2), & 
\mbox{if}\ \  s_2 \leq s_1,\ \gamma >0, \beta > 0.\\
\end{array} \right.
\ees
\end{theorem}

\section {Simulation Studies.}
\label{sec:simulations}
\setcounter{equation}{0}

In order to study   finite sample properties of the proposed estimation procedure, we carried out a 
limited simulation study. For each test function $f(t,\bx)$ and a kernel $g(t)$,
we obtained exact values of $q(t,\bx)$ in the equation \fr{eq1} by integration.
We considered  $n$ equally spaced points $t_k = T k/n$, $k=1, \cdots, n$, on the time interval $[0;T]$. 
We created a uniform grid $\lfi x_{1,i}, x_{2,j}\rfi$ on $[0,1]\times [0,1]$ 
with $i=1, \cdots, n_1$ and $j=1, \cdots, n_2$, and obtained the three-dimensional array   
$q_{i,k,j} = q(x_{1,i}, x_{2,j}, t_k)$. After fixing the   Signal-to-Noise Ratio (SNR), we evaluated the value of
$\sig$ as  $\sig = n^{-1/2}\, \std(q)/\SNR$, where $\std(q)$ is the standard deviation of the tensor  with values
$q_{i,k,j}$  reshaped as a vector. 
Finally, we obtained a sample $Y_{i,j,k}$ of the left-hand side of 
the equation \fr{eq1} by adding independent Gaussian $\NN(0,\sig^2)$ noise to each value $q_{i,k,j}$,
 $i=1, \cdots, n_1$, $j=1, \cdots, n_2$, $k=1, \cdots, n$. 

\begin{table}  
\begin{center}
\begin{tabular}{|l| c |c |c|c| c|    }
%
\hline
  Function   &  $\std(f)$ & $\|f\|$  &    SNR=3  &     SNR=5    & SNR=7   \\
\hline
 \hline
$f_1(t, \bx)$  & 0.0025 & 0.5084         & 0.1107  (0.0110)   & 0.0694  (0.0066)  &  0.0511 (0.0049)     \\
\hline
$f_2(t, \bx)$  & 0.3334 & 61.8367        & 0.1224   (0.0100)  &   0.0761  (0.0071) &  0.0567  (0.0051)  \\
\hline
$f_3(t, \bx)$   & 0.3342 & 62.0261      &  0.1107   (0.0112)   &   0.0680  (0.0068) &  0.0511  (0.0048)  \\
\hline
$f_4(t, \bx)$   & 0.3366 & 62,6863      & 0.1080   (0.0117)   &   0.0690   (0.0058) &  0.0519  (0.0046)  \\
 \hline
 \hline
\end{tabular}
\end{center}
\caption{ The standard deviations, the norms and the  average values of the relative errors $\Del(\widehat{f})$ 
(with the standard errors of the means in  parentheses)
evaluated over 100 simulation runs for the four test functions. 
The test functions are defined in formula \fr{test_fun}.
}
\end{table}

We constructed a system of $M$   Laguerre functions of the form \fr{eq:Laguerre_func}.
For each time point $k=1,\cdots, n$, we found the matrix of wavelet coefficients using the  Daubechies 6 wavelets
and constructed estimators $\widehat{\sig}_k$, $k=1,\cdots, n$, of $\sigma$ as the standard deviations of the wavelet coefficients at
the highest resolution level. Subsequently, we  obtained  $\widehat{\sig}$ as the average of $\widehat{\sig}_k$, $k=1,\cdots, n$.
Finally, for each of the indices $\bom \in \Om(J_1,J_2)$, we evaluated the sample wavelet-Laguerre coefficients $\widehat{\theta}_{l; \bom}$, 
$l = 0, \cdots, M-1$, as   solutions of the linear regression problems.  

Next, for each $l = 0, \cdots, M-1$, we derived the threshold  $\lambda_{l, \widehat{\eps}}$ 
of the form \fr{Thres} with $\widehat{\eps}  = T \widehat{\sig}/\sqrt{n}$ and $n = n_0 n_1 n_2$,
and obtained the thresholded estimators $\widehat{\theta}_{l; \bom}\, \II \left(| \widehat{\theta}_{l; \bom}|  > \lambda_{l, \widehat{\eps}}  \right)$ 
of the coefficients $\theta_{l; \bom}$, $l = 0, \cdots, M-1$, $\bom \in \Om(J_1, J_2)$. 
Finally we constructed the estimator $\widehat{f}$ of the form \fr{fhat}
by the Laguerre reconstruction and the subsequent inverse wavelet transforms.

In our simulations, we used $n_1= n_2 = n = 32$, $M=8$  and $T=5$. We chose 
$g(x) =  \exp(-x/2)$ and carried out simulations with the following test functions
\begin{align}
f_1 (t, \bx) & =  t\, e^{-t}  (x_1-0.5)^2)\, (x_2-0.5)^2, \nonumber \\
f_2 (t, \bx) & = e^{-t/2} \, \cos(2 \pi x_1 x_2),  \label{test_fun} \\
f_3 (t, \bx) & = t\, e^{-t}  (x_1-0.5)^2)\, (x_2-0.5)^2 + e^{-t/2} \, \cos(2 \pi x_1 x_2),\nonumber \\
f_4 (t, \bx) & = e^{-t/2} \, \cos(2 \pi x_1 x_2) +  (x_1-0.5)^2 \, (x_2-0.5)^2. \nonumber 
\end{align}
We also considered three noise scenarios: SNR = 3 (high noise level), SNR = 5 (medium noise level) 
and SNR = 7 (low noise level).  In order for the values of the errors of our estimators to be independent of the 
norms of the test functions, we evaluated the average relative error  as the average $L^2$-norm  
of the difference between $f$ and its estimator divided by the norm of $f$:
\bes
\Del(\widehat{f}) = \|\widehat{f} - f \|/\|f\|.  
\ees
 Table 1  reports the mean values of those errors over 100 simulation runs 
(with the standard errors of the means presented in  parentheses) 
for the four  test functions and the three noise levels.
The errors are reported together with the  standard deviations and the norms of each of the functions.

Table 1 confirms that our method allows to solve the functional deconvolution problem with high accuracy. 
As it is expected, the precision of estimation improves when $\SNR$ grows and $\sig$ declines. 
Note also that  reporting the relative errors for each of the test functions and arranging them in accordance with the  
SNR values  allows us, in some way, to characterize precision of the method rather than the complexity of 
the recovery of a particular test function. Indeed, the relative errors of  estimators of all four test functions  
are similar to each other in spite of variations in their norms and standard deviations.

\section {Real Data Example.}
\label{sec:real_data}
\setcounter{equation}{0}

As an application of the proposed technique we studied the recovery of the unknown function 
$f(t, \bx) = \beta (1 - F(t, \bx))$ in the   equation \fr{complete_DCE} on the basis 
of the DCE-CT (Computerized Tomography)   images of a participant 
of the REMISCAN cohort study \cite{remiscan}  who underwent anti-angiogenic treatment for  renal cancer.
The data consist  of the arterial images and images of the area of interest (AOI) at 
37 time points over approximately 4.6 minute interval. The first 15 time points (approximately the first 
30 seconds) correspond to the time period before the contrast agent reached the aorta and the AOI  
(so  $\del = 0$ in equation \fr{complete_DCE}). 
We   used those data points for the evaluation of the base intensity. 

Since the  images of the aorta  are extremely noisy, we evaluated the average values of the 
grey level intensity at each time point and then used Laguerre functions smoothing in order to obtain
the values of the Arterial Input Function $\AIF(t)$. The images of AOI contain $49 \times 38$ pixels.
Since our technique is based on  periodic wavelets and hence application of the method to a non-periodic function
is likely to produce Gibbs effects, we cut the images to the size  of $32 \times 32$ pixels.
Furthermore, in order to achieve  periodicity, we  obtained symmetric versions of  the  images (reflecting the images over the two sides)
and applied our methodology to the resulting spatially periodic functions.  Consequently, the estimator 
obtained by the technique   is spatially symmetric, so we record only the original part as the estimator  $\widehat{f}$.
Figure 1   shows the averages of the aorta intensities at each time point and its de-noised version that was used as $\AIF(t)$.
Figure 2 presents the values of $\widehat{f}$ at 34 seconds (corresponds to the first time point when the contrast agent 
reaches the AOI), 95 seconds (the 12-th time point) and 275 seconds (the last time point).  

\begin{figure} 
\[\includegraphics[height=6.0cm]{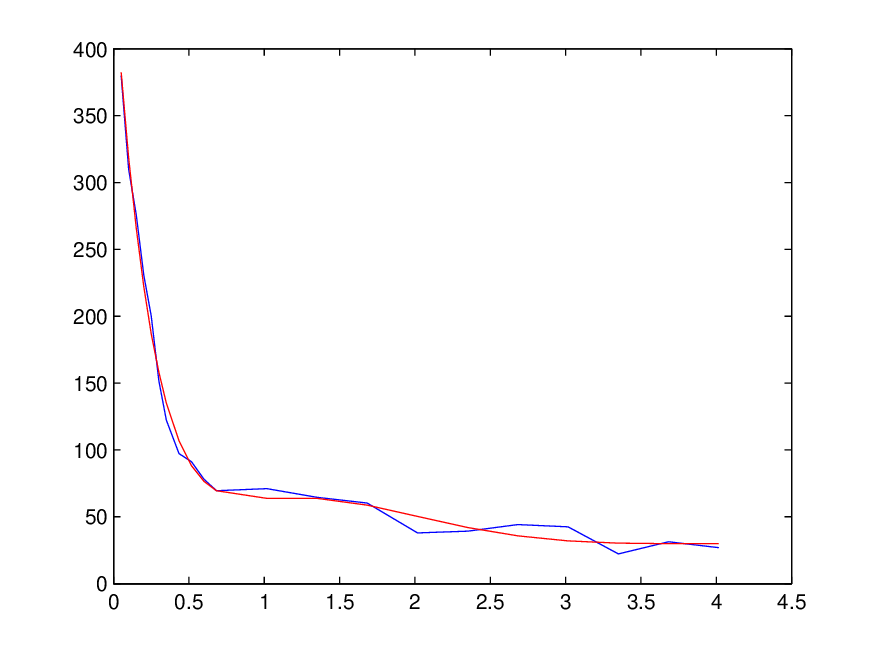} \hspace{2mm} 
 \hspace{2mm} \includegraphics[height=6.0cm]{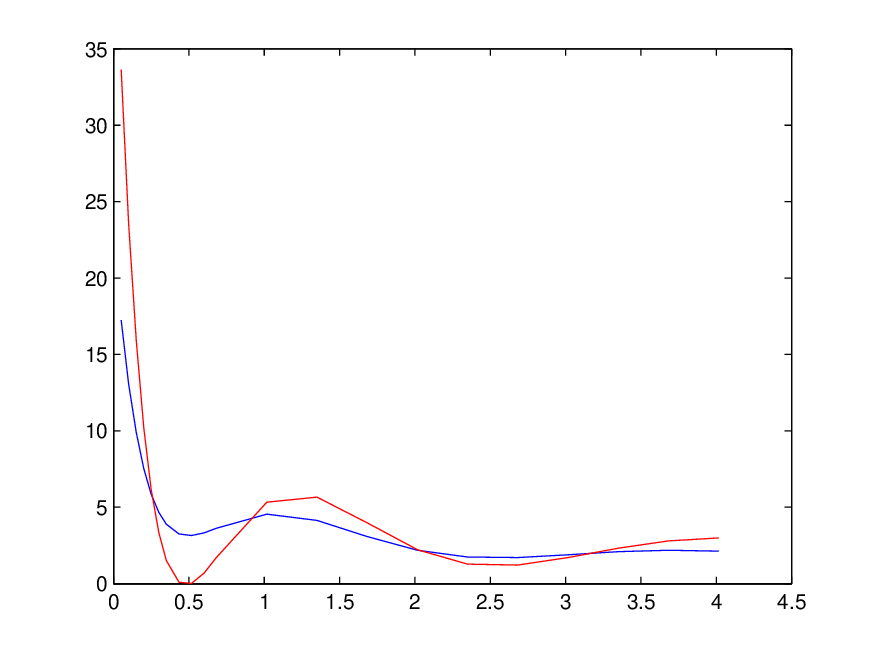} \]
\caption{Left: the averages of the aorta intensities (blue) and the estimated  
Arterial Input Function $\AIF(t)$ (red). Right: two   curves for distinct spatial locations.
\label{fig1}}
\end{figure}

\begin{figure} 
\[\includegraphics[height=4.0cm]{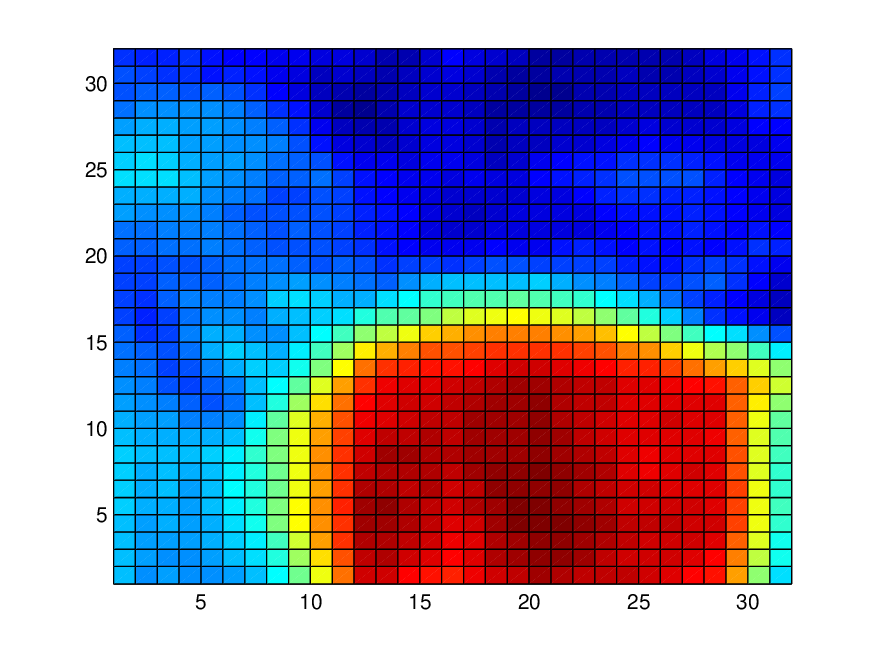} \hspace{2mm} 
\includegraphics[height=4.0cm]{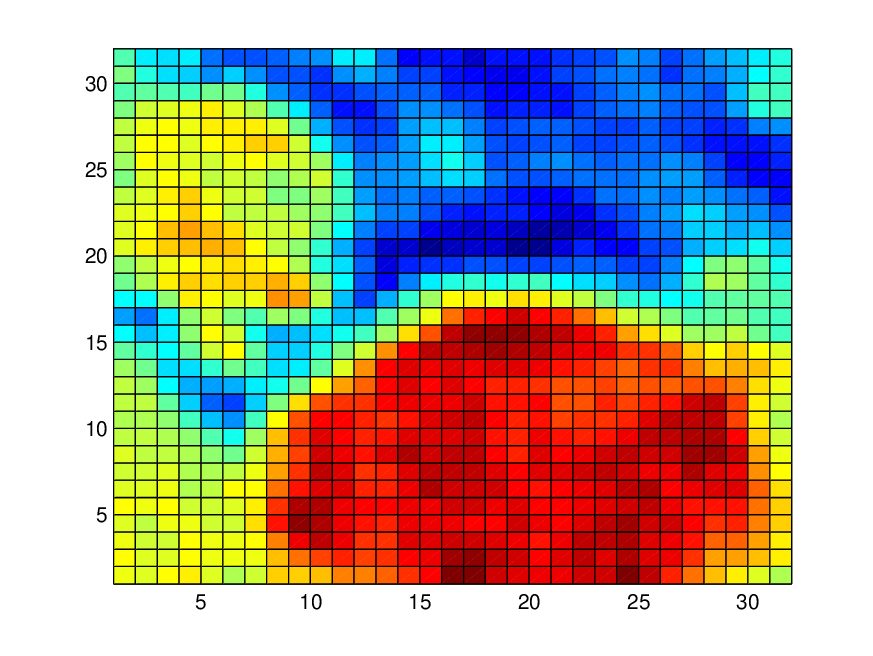} \hspace{2mm}
 \includegraphics[height=4.0cm]{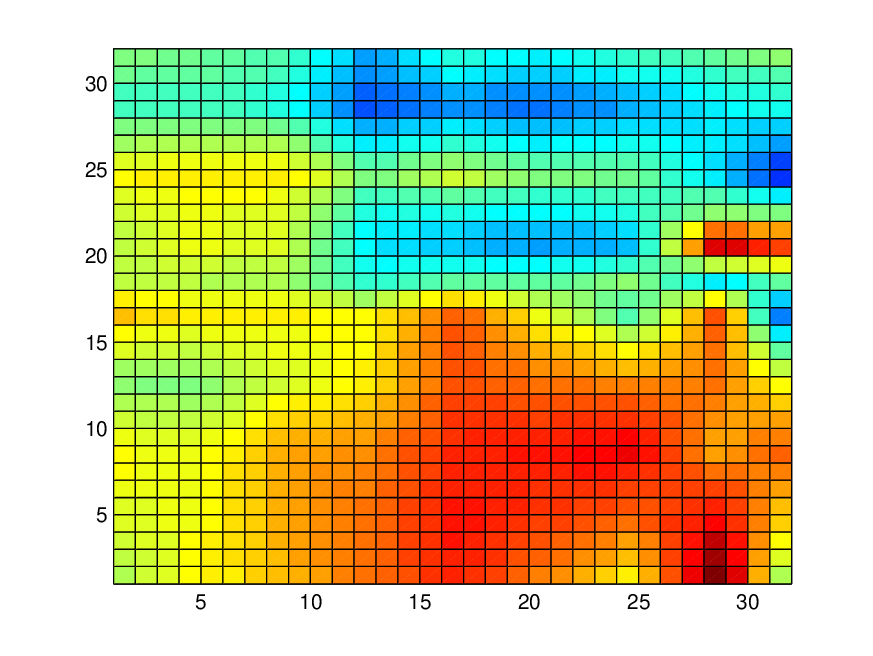}  \]
\caption{The values of $\widehat{f}$ at 34 seconds (corresponds to the first time point), 95 seconds
(the 12th time point) and 275 seconds (the last time point).  
\label{fig2}}
\end{figure}

\section*{Acknowledgments}

Marianna Pensky  and Rasika Rajapakshage were  partially supported by the National Science Foundation
(NSF), grants   DMS-1407475 and DMS-1712977.



\section {Proofs. }
\label{sec:proofs}
\setcounter{equation}{0}
 
\subsection{Proof of the lower bounds for the risk.}

In order to prove Theorem \ref{th:lowerbds}, we use Lemma A1 
of Bunea {\it et al.}~(2007), which we will reformulate for the squared risk case.

\begin{lemma} \label{lem:Bunea} 
Let $\Te$ be a set of functions of cardinality $\card(\Te)\geq 2$ such that\\
(i) $\|f-g\|^2 \geq 4\delta^2, \ for\  f, g \in \Te, \ f \neq g, $\\
(ii) the Kullback divergences $K(P_f, P_g)$ between the measures $P_f$ and $P_g$ 
satisfy the inequality $K(P_f, P_g) \leq \log(\card(\Te))/16,\ for\ f,\ g \in \Te$.\\
Then, for some absolute positive constant $C_1$, one has 
$$
 \inf_{f_n}\sup_{f\in \Te} \EE_f\|f_n-f\|^2 \geq C_1 \delta^2,
$$
where $\inf_{f_n}$ denotes the infimum over all estimators.
\end{lemma}

In order to obtain lower bounds, we introduce a  triangular  Toeplitz matrix
associated with Laurent series $(1-z)^{-r}$ (see Section~\ref{sec:Toeplitz}  for more detailed explanations) and denote   by 
$Q^{(L)} = T_L\lkr (1-z)^{-r} \rkr$  its reduction to the set of indices $0 \leq l \leq L-1$.
Following Vareschi (2013), consider   function 
\be \label{ht} 
h(t) = \sum_{l=0}^\infty h_l \ph_l(t) \quad \mbox{with} \ 
h_l = \frac{(-1)^l}{\log(l\vee e)}\, {-1/2 \choose l} = 
\frac{\Ga\lkr\frac{1}{2} \rkr\, \Ga\lkr\frac{1}{2} + l \rkr}{\Ga(l+1)\, \log(l\vee e)}.  
\ee
 Denote $\bte^{(L)} = (\te_0, \cdots, \te_{L-1})^T = Q^{(L)} \bh^{(L)}$ where 
$\bh_L$ is the vector of the first $L$ coefficients of function $h$ in \fr{ht}.
In what follows we shall use Lemma 6.5 of Vareschi (2013) 
that was in the original version of the paper posted on ArXiv but did not make it 
to the published version  of Vareschi (2015).

\begin{lemma} \label{lem:vareschi} 
Let $h(t)$ be as  defined in \fr{ht} and  $\bte^{(L)} =   Q^{(L)} \bh^{(L)}$ where $Q^{(L)} = T_L\lkr (1-z)^{-r} \rkr$ 
and $\bh^{(L)}$ are reductions of the infinite-dimensional Toeplitz matrix $T\lkr (1-z)^{-r} \rkr$ and  vector $\bh$
of coefficients of $h(t)$ to the set of indices $0 \leq l \leq L-1$. Then, $h(t)$ is square integrable and 
there exist positive constants $C_{r1}$and $C_{r2}$ that depend on $r$ only such that for all $r \geq 1$ and any $l \geq 0$ one has
\be \label{eq:vareschi}
C_{r1}\, \frac{(l \vee 1)^{r-1/2}}{\log(l \vee e)} \leq \te_l \leq   
C_{r2}\,  (l \vee 1)^{r-1/2}. 
\ee
\end{lemma}

\noindent
Let $\vart$ be a matrix with components $\vart_{k_1, k_2}=\left \{-1, 1 \right\}$, $k_1=0, 1, \cdots, 2^{j_1}-1$, 
$k_2=0, 1, \cdots, 2^{j_2}-1$. Denote the set of all possible values of $\vart$ by 
$\Te$ and let  functions $f_{L,j_1, j_2}$ be of the form 
\begin{align}   \label{f_j}
& f_{L,j_1, j_2}(t, x_1, x_2)=\rho \  q_L(t)\ p_{j_1, j_2}(x_1, x_2), \\
& q_L(t) = \sum^{L-1}_{l=0} \theta_l\, \ph_l(t), \ \ 
p_{j_1, j_2}(x_1, x_2)= \sum^{2^{j_1}-1}_{k_1=0}\  \sum^{2^{j_2}-1}_{k_2=0}  \vart_{k_1, k_2}\, 
\, \psi_{j_1, k_1}(x_1)\, \psi_{j_2, k_2}(x_2), \label{qLg}
\end{align} 
where  $\bte^{(L)}$ is the vector with components $\te_l$, $l=0, \cdots, L-1$ where $\bte^{(L)} =Q^{(L)} \bh^{(L)}$ 
and $Q^{(L)}$ and $\bh^{(L)}$ are defined above.   
Since $f_{L,j_1, j_2} \in {\cal B}_{\gamma, \beta}^{s_1, s_2, s_3}(A)$, Lemma \ref{lem:vareschi} implies that   
one can choose
\be \label{rho_value} 
\rho^2  = C_r A^2
 2^{-2j_1(s_1+\frac{1}{2})- 2j_2(s_2+\frac{1}{2})}\, (L \vee 1)^{-2(r+s_3)}\, \exp \lfi - 2 \gamma L^\beta \rfi,
\ee
where $0 < C_r  \leq C_{r2}^2 /2r$.
If $\tilde{f}_{L,j_1, j_2}$ is of the form \fr{f_j} but with $\tilde{\vart}_{k_1, k_2}  \in \Te$ instead of 
$\vart_{k_1, k_2}$, then, by Lemma \ref{lem:vareschi}, the $L^2$-norm of the difference is of the form 
$$
\|\tilde{f}_{L, j_1, j_2} - f_{L, j_1, j_2} \|_2^2 =   \rho^2  \lkr \sum^{L-1}_{l=0}\, \te_l^2 \rkr \  
\lkr \sum^{2^{j_1}-1}_{k_1=0}\, \sum^{2^{j_2}-1}_{k_2=0}\, 
 \II\left(\tilde{\vart}_{k_1, k_2} \neq {\vart}_{k_1, k_2} \right) \rkr \geq 
\frac{C_{r1}^2\rho^2  H\left( \tilde{\vart}, \vart \right) (L \vee 1)^{2r}}{2r\, [\log (L \vee e)]^2}.
$$
Here $H\left( \tilde{\vart}, \vart \right)$ is the Hamming distance between the binary sequences $\vect(\vart)$  
and $\vect(\tilde{\vart})$ where $\vect(\vart)$ is a vectorized version of matrix $\vart$.

Observe that   matrix $\vart$ has $\aleph=2^{j_1 + j_2}$ components, and hence, $\card(\Te)= 2^{\aleph}$. 
In order to find a lower bound for $H\left( \tilde{\vart}, \vart \right)$, we apply the Varshamov-Gilbert lemma 
which states that one can choose a subset ${\Te_1}$ of ${\Te}$, of cardinality of at least $2^{\aleph/8}$, and  
such that $H\left( \tilde{\vart}, \vart \right) \geq \frac{\aleph}{8}$ for any $\vart, \tilde{\vart} \in \Te_1$. 
Hence, for any $\vart, \tilde{\vart} \in \Te_1$, one has the following expression for 
$\delta^2$   defined  in Lemma~\ref{lem:Bunea}:
\be   \label{del}
\|\tilde{f}_{L,  j_1, j_2} - f_{L,   j_1, j_2} \|^2 \geq 
\frac{C_{r1}^2 \rho^2 2^{j_1+j_2} (L \vee 1)^{2r}}{16r\, [\log (L \vee e)]^2} =  4 \delta^2.
\ee 
Let $P_f$ be the distribution of the process $\left\{ f*g(t, {\bf x})+ \varepsilon dW(t, {\bf x}), (t, {\bf x}) \in U\right\}$ when $f$ is true, where $W(t, {\bf x})$ is a Wiener process. Then, since $\left| \tilde{\vart}_{l, k, k'}-\vart_{l, k, k'}\right| \leq 2$, and
{due to} the multiparameter Girsanov formula (see, e.g., Dozzi (1989), p. 89),
  \fr{f_j} and \fr{qLg},  the Kullback divergence can be bounded as 
\beqn     \label{KLbk}
 K(P_f, P_{\tilde{f}})  & = & \EE_{P_f} \left[ \log \left( P_f/P_{\tilde{f}}\right)\right]
 =-\EE_{P_f}\left[\varepsilon^{-1} \int_U ((f-\tilde{f})*g)(t, {\bf x})dW(t, {\bf x})\right] \nonumber\\
 &+& \left(2 \varepsilon^2 \right)^{-1} \int_U\left((f-\tilde{f})*g\right)^2(t, {\bf x})dtd{\bf x} \nonumber\\
   & = & \left(2 \varepsilon^2 \right)^{-1}  \|( \tilde{f}-f )*g\|_2^2 \nonumber\\
& \leq & 2 \varepsilon^{-2}\, \rho^2  2^{j_1+j_2} \|q_L*g\|_2^2
= 2 \varepsilon^{-2}\,  \rho^2  2^{j_1+j_2} \| G^{(L)} \bte^{(L)} \|^2_2,    
\eeqn 
where matrix $G^{(L)}$ and vector $\bte^{(L)}$ are defined in \fr{Toeplitz_matr} and Lemma \ref{lem:vareschi},
respectively. By Lemma~\ref{lem:comte_lem3_4} in section \ref{sec:Toeplitz}, and under Assumptions {\bf A1} and {\bf A2}, one 
obtains that $G^{(L)}  =  T_L((1-z)^r v(z))$ and  
$\|T_L(v(z))\|^2 = \lam_{\max} [T_L^T(v(z)) T_L(v(z))] < \|v\|_{circ}^2 < \infty$. 
Therefore, $G^{(L)} \bte^{(L)} = G^{(L)} Q^{(L)} \bh^{(L)}=  T_L((1-z)^r v(z)) T_L((1-z)^{-r}) \bh^{(L)}$
and
\be \label{GLte}
\|G^{(L)} \bte^{(L)}\|_2^2 = \|T_L(v(z))  \bh^{(L)}\|_2^2 \leq \|T_L(v(z))\|^2 \|\bh^{(L)}\|_2^2 
\leq  \|v\|_{circ}^2\, \|h\|_2^2 <\infty,
\ee
where $\|h\|_2^2$ is the $L^2$-norm of the function $h(t)$ and $\|h\|_2 < \infty$ due 
to Lemma~\ref{lem:vareschi}. Combination of \fr{KLbk} and \fr{GLte}
yields $K(P_f, P_{\tilde{f}})  \leq   \tilde{C} \eps^{-2} \rho^2 2^{j_1+j_2}/16$ where 
$\tilde{C} = 32 \|v\|_{circ}^2 \|h\|_2^2$.
Application of  Lemma \ref{lem:Bunea} requires the constraint 
$$
K(P_f, P_{\tilde{f}})  \leq  \log(\card(\Te))/16= \log(2)2^{j_1 + j_2}/16.
$$ 
Therefore, one can choose $\rho^2 = \eps^2/\tilde{C}$, so that,
by Lemma \ref{lem:Bunea} for some $C_1 >0$ one has   
\be \label{eq:low_bound}
\inf_{f_n}\sup_{f\in \Te} \EE_f\|f_n-f\|^2 \geq C_1 
 \eps^2 2^{j_1+j_2} \,(L \vee 1)^{2r} \,[\log (L \vee e)]^{-2},
\ee
where $L$, $j_1$ and $j_2$ are such that 
\be \label{ineq_cond}
2^{2j_1(s_1+\frac{1}{2})+ 2j_2(s_2+\frac{1}{2})}\, (L \vee 1)^{2(r+s_3)}\, \exp \lfi   2 \gamma L^\beta \rfi
= C_2 A^2 \eps^{-2},
\ee
with $C_2 =  C_r \tilde{C}/\log(2)$.
Thus, one needs to choose   $j_1$, $j_2$ and $L$ that maximize $2^{j_1+j_2} \,(L \vee 1)^{2r} \,[\log (L \vee e)]^{-2}$
subject to condition \fr{ineq_cond}. Denote
\be
\tau_{\eps}=\log({A^2}{\varepsilon^{-2}}).
\ee

It is easy to check that the solution of the above linear constraint optimization problem is of the form $\left\{ j_1, j_2, L\right\}= 
\left\{0, 0, \left[ A^2 \varepsilon^{-2} \right]^{\frac{1}{2s_3 + 2r}} \right\}$ if $s_3 \leq \min \{2rs_1, 2rs_2\}$ and $\gamma=\beta=0$, 
 $\left\{ j_1, j_2, L\right\}= \left\{0, \left(\log(2) \right)^{-1}\left(2s_2+1 \right)^{-1}\tau_{\varepsilon}, e \right\}$ 
if $s_1 \geq s_2$, $s_3 \geq 2rs_2-2s_2-1$ and $\gamma=\beta=0$, 
$\left\{ j_1, j_2, L\right\}= \left\{ \left(\log(2) \right)^{-1}\left(2s_1+1 \right)^{-1}\tau_{\varepsilon}, 0, e \right\}$ 
if $s_1 \leq s_2$ and $s_3 \geq 2rs_1-2s_1-1$ and $\gamma=\beta=0$. $\left\{ j_1, j_2, L\right\}= \left\{0, \left(\log(2) \right)^{-1}\left(2s_2+1 \right)^{-1}\tau_{\varepsilon}, e \right\}$ 
if $s_1 \geq s_2$ and $\gamma>0, \beta>0$, and $\left\{ j_1, j_2, L\right\}= \left\{ \left(\log(2) \right)^{-1}\left(2s_1+1 \right)^{-1}\tau_{\varepsilon}, 0, e \right\}$ 
if $s_1 \leq s_2$ and $\gamma>0, \beta>0$.
By noting that 
\be \label{}
\frac{s_3}{s_3+r} \leq \min\left\{ \frac{2s_2}{2s_2+1}, \frac{2s_1}{2s_1+1}\right\}, \ \ if \ s_3 \leq \min\{2rs_1, 2rs_2 \}, \ \ \ \gamma=\beta=0,\\
\ee
\be \label{}
\frac{2s_1}{2s_1+1}\leq \min\left\{ \frac{2s_2}{2s_2+1}, \frac{s_3}{s_3+r}\right\}, \ \ if \ s_1 \leq \min\{{s_3}/{2r}, s_2 \}, \ \ \ \gamma=\beta=0,\\
\ee
\be \label{}
\frac{2s_2}{2s_2+1}\leq \min\left\{ \frac{2s_1}{2s_1+1}, \frac{s_3}{s_3+r}\right\}, \ \ if \ s_2 \leq \min\{{s_3}/{2r}, s_1 \}, \ \ \ \gamma=\beta=0,\\
\ee
and 
\be \label{}
\frac{2s_1}{2s_1+1}\leq \frac{2s_2}{2s_2+1}, \ \ if \ s_1 \leq s_2,\ \ \ \gamma >0, \beta > 0,
\ee
we then choose the highest lower bounds in \fr{eq:low_bound}. This completes the proof of the theorem.

\subsection{Proof of the upper bounds for the risk.}
{\bf The proof of Lemma \ref{lem:VarDev}.  }Denote the quantities $\widehat{\theta}_{l, \bom}-\theta_{l, \bom}$ by $\aleph_{l, \bom}$, and notice that $\aleph_{l, \bom}=\widehat{\theta}_{l, \bom}-\theta_{l, \bom}=\varepsilon {\bf e_l}^T\left({\bf G}^{(l)}\right)^{-1}\xi^{(l)}$, where $\xi^{(l)}$ is the $l$-dimensional Gaussian vector such that $\xi^{(l)} \sim N(0, I_l)$, and ${\bf e_l}$ is the $l^{th}$ standard basis vector of dimension $l$. Also, note that $\varepsilon {\bf e}_l^T({\bf G}^{(l)})^{-1}\xi^{(l)}= \varepsilon\sum^{l-1}_{k=0}v_k\xi_k$, where $v_k$ is defined in Lemma 1. Then, by \fr{elem_bounds}, the variance of $\aleph_{l, \bom}$ is
\be
\EE\left[\aleph_{l, \bom}\right]^2= \varepsilon^2 \sum^{l-1}_{k=0}v_k^2\leq C_{v_2}\varepsilon^2 l^{2r-1}.
\ee 
Now, for the fourth moment of $\aleph_{l, \bom}$, and using properties of Gaussian random variables, one has 
\beqns
\EE\left[\aleph_{l, \bom}\right]^4&=&\varepsilon^4\EE\left[\sum^{l-1}_{k=0}v_k\xi_k\right]^4\nonumber\\
&=& \varepsilon^4\left[ \sum^{l-1}_{k=0}v_k^4\EE(\xi_k^4) + 3\sum^{l-1}_{k_1, k_2=0, k_1\neq k_2}v^2_{k_1}v^2_{k_2}\right]\nonumber\\
&=& 3\varepsilon^4\left[ \sum^{l-1}_{k=0}v^2_k\right]^2 \leq 3\varepsilon^4\left[C_{v_2}l^{2r-1}\right]^2.
\eeqns
This completes the proof of $\fr{mom4}$. In order to prove formula $\fr{Largdev}$, recall that $\aleph_{l, \bom} \sim N(0, \varepsilon^2\sum^{l-1}_{k=0}v_k^2)$. Therefore, by the Gaussian tail probability inequality, one obtains
\be
\Pr\left( |\aleph_{l, \bom}| > \sqrt{2\tau \ln(\varepsilon^{-1})}\varepsilon \sqrt{\sum^{l-1}_{k=0}v_k^2}\right) \leq  \left[\tau\pi \ln(\varepsilon^{-1})\right]^{-1/2}\varepsilon^{\tau}.
\ee
Now, since
\be
\sum^{l-1}_{k=0}v_k^2 \leq C_{v_2}l^{2r-1}\leq \frac{C_{v_2}}{C_{G_1}}l^{-1}C_{G_1}l^{2r}\leq \frac{C_{v_2}}{C_{G_1}}l^{-1}\|({\bf G}^{(l)})^{-1}\|,
\ee
 $(4.5)$ follows, provided $\nu \geq \tau \frac{C_{v_2}}{C_{G_1}}$. \\
 {\bf The proof of Theorem \ref{th:upperbds}.  }Denote 
\beqn \label{d}
\mu=\left\{ \begin{array}{ll}\min\left\{\frac{s_3}{s_3+r}, \frac{2s_2}{2s_2+1}, \frac{2s_1}{2s_1+1}\right\}, \ \ if  \ \ \ \gamma=\beta=0,\\
\min\left\{ \frac{2s_2}{2s_2+1}, \frac{2s_1}{2s_1+1}\right\}, \ \ \ \ if  \ \ \ \ \ \ \gamma >0, \beta > 0.\\
\end{array} \right.
\eeqn
\beqn  \label{Lev:chi}
\chi_{\varepsilon, A}= \left[{A^{-2}} {\varepsilon^{2}}\log(1/\varepsilon)\right],
\eeqn

\beqn  \label{Lev:J0}
2^{j_{10}}= \left[\chi_{\varepsilon, A}\right]^{-\frac{\mu}{2s_1}},\ \ \ 2^{j_{20}}= \left[\chi_{\varepsilon, A}\right]^{-\frac{\mu}{2s_2}},
\eeqn
and 
\beqn  \label{Lev:M0}
M_0= \left\{ \begin{array}{ll} \left[\chi_{\varepsilon, A}\right]^{-\frac{\mu}{2s_3}}& \mbox{if}\ \ \gamma = \beta = 0\\
\left[\frac{\log(1/\varepsilon)}{\gamma}\right]^{1/\beta} & \mbox{if}\ \ \gamma>0, \beta > 0,\\
\end{array} \right.
\eeqn
and notice that with the choices of $J_1$, $J_2$ and $M$ given by \fr{Lev:J}, the estimation error can be decomposed into the sum of three components as follows
\beqn \label{eseroll}
\EE \| \widehat{f}_n-f\|^2 \leq \sum_{\omega}\sum^{\infty}_{l=0}\EE \| \widehat{\theta}_{l: \bom}\II \left(| \widehat{\theta}_{l:\bom}|  > \lambda_{l, \varepsilon}   \right)-\theta_{l: \bom} \|^2\leq R_1+ R_2 + R_3, 
\eeqn
where 
\beqns
R_1&=& \sum^{J_1-1}_{j_1=0}\sum^{J_2-1}_{j_2=0}\sum^{M-1}_{l=0}\sum_{k, k'}\EE \left[\left| \widehat{\theta}_{l: \bom}-\theta_{l: \bom}\right|^2 \II \left(  \left| \widehat{\theta}_{l: \bom}  \right| >\lambda_{l, \varepsilon} \right)  \right], \nonumber\\
R_2&=& \sum^{J_1-1}_{j_1=0}\sum^{J_2-1}_{j_2=0}\sum^{M-1}_{l=0} \sum_{k, k'}\left| {\theta}_{l: \bom}  \right|^2 \Pr \left( \left| \widehat{\theta}_{l: \bom}  \right| < \lambda_{l, \varepsilon} \right),\\
R_3&=&  \left(\sum^{\infty}_{j_1=J_1}\sum^{\infty}_{j_2=J_2}\sum^{\infty}_{l=M}+\sum^{J_1-1}_{j_1=0}\sum^{\infty}_{j_2=J_2}\sum^{\infty}_{l=M}+ \sum^{\infty}_{j_1=J_1}\sum^{J_2-1}_{j_2=0}\sum^{\infty}_{l=M}+ \sum^{\infty}_{j_1=J_1}\sum^{\infty}_{j_2=J_2}\sum^{M-1}_{l=0}\cdots \right)\sum_{k, k'}\left| \theta_{l: \bom}  \right|^2.
\eeqns
For $R_3$, one uses assumption \fr{LagSob} to obtain, 
\beqn  \label{R_3}
R_3&=& O \left( \left(\sum^{J_1-1}_{j_1=0}\sum^{J_2-1}_{j_2=0}\sum^{\infty}_{l=M}+\sum^{J_1-1}_{j_1=0}\sum^{\infty}_{j_2=J_2}\sum^{M}_{l=1}+ \sum^{\infty}_{j_1=J_1}\sum^{J_2-1}_{j_2=0}\sum^{M}_{l=1} \right)A^22^{-2j_1s_1 -2j_2s_2}l^{-2s_3}\exp \{-2\gamma l^{\beta} \}\right)\nonumber\\
&=& O \left( A^2 2^{-2J_1s_1} + A^2 2^{-2J_2s_2} +A^2 M^{-2s_3 }\exp \{-2\gamma M^{\beta} \} \right).\nonumber\\
\eeqn
If $\gamma=\beta=0$, then since $M\asymp \left[\varepsilon^2\right]^{-1/2r}$, $R_3$ becomes
\beqn  \label{R'_3}
R_3&=& O \left( A^2\left[ A^{-2}\varepsilon^2 \right]^{2s_1}+ A^2\left[ A^{-2}\varepsilon^2 \right]^{2s_2} + A^2\left[ A^{-2}\varepsilon^2 \right]^{\frac{2s_3}{2r}}  \right)\nonumber\\
&=& O \left( A^2 \left[ \chi_{\varepsilon, A}\right]^{\mu} \right).
\eeqn
If $\gamma>0, \beta>0$, then 
\beqn  \label{R''_3}
R_3&=& O \left( A^2\left[ A^{-2}\varepsilon^2 \right]^{2s_1}+ A^2\left[ A^{-2}\varepsilon^2 \right]^{2s_2}   \right)\nonumber\\
&=& O \left( A^2 \left[ \chi_{\varepsilon, A}\right]^{\min \left \{ \frac{2s_2}{2s_2+1}, \frac{2s_1}{2s_1+1}\right \} } \right).
\eeqn

To evaluate the remaining two terms, notice that both $R_1$ and $R_2$ can be partitioned into the sum of two error terms as follows
\beqn  \label{R_2}
R_1\leq R_{11} + R_{12}, \ \ \ \ R_2 \leq R_{21} + R_{22},
\eeqn
where 
\beqn  
R_{11}&=&\sum^{J_1-1}_{j_1=0}\sum^{J_2-1}_{j_2=0}\sum^{M}_{l=1}\sum_{k, k'}\EE \left[\left| \widehat{\theta}_{l: \bom}-\theta_{l: \bom}\right|^2 \II \left(  \left| \widehat{\theta}_{l: \bom}-\theta_{l: \bom} \right| >  \frac{1}{2}\lambda_{l;\varepsilon} \right)  \right], \label{r21}\\
R_{12}&=& \sum^{J_1-1}_{j_1=0}\sum^{J_2-1}_{j_2=0}\sum^{M}_{l=1}\sum_{k, k'}\EE \left[\left| \widehat{\theta}_{l: \bom}-\theta_{l: \bom}\right|^2 \II \left(  \left|  \theta_{l: \bom} \right| >  \frac{1}{2}\lambda_{l;\varepsilon} \right)  \right],\label{r22}\\
R_{21}&=& \sum^{J_1-1}_{j_1=0}\sum^{J_2-1}_{j_2=0}\sum^{M}_{l=1}\sum_{k, k'} \left| \theta_{l: \bom}  \right|^2 \Pr \left( \left|  \widehat{\theta}_{l: \bom}-\theta_{l: \bom}  \right| > \frac{1}{2} \lambda_{l;\varepsilon} \right),\label{r31}\\
R_{22}&=& \sum^{J_1-1}_{j_1=0}\sum^{J_2-1}_{j_2=0}\sum^{M}_{l=1}\sum_{k, k'} \left| \theta_{l: \bom}  \right|^2\II \left(  \left|  \theta_{l: \bom} \right| <  \frac{3}{2}\lambda_{l;\varepsilon} \right).\label{r32}
\eeqn
Combining \fr{r21} and \fr{r31} and applying Cauchy-Schwarz inequality, Lemma \ref{lem:VarDev} and the fact that $M\asymp \left[\varepsilon^2\right]^{-1/2r}$, yields
\beqns
R_{11} + R_{21}&=& O \left(\sum^{J_1-1}_{j_1=0}\sum^{J_2-1}_{j_2=0}\sum^{M}_{l=1} \left(2^{j_1+j_2}\varepsilon^2 l^{2r-1}\varepsilon^{\tau/2}+ \varepsilon^{\tau}  \sum_{k, k'} \left| \theta_{l: \bom}  \right|^2 \right)\right)\\
&=& O  \left( {\varepsilon^{2}}2^{J_1+J_2}M^{2r}\left( {\varepsilon^{2}}\right)^{\frac{\tau}{4}} + A^2\varepsilon^{\tau} \right)\\
&=& O  \left(A^4\left( {\varepsilon^{2}}\right)^{\frac{\tau}{4}-2} + A^2\varepsilon^{\tau} \right).
\eeqns
Hence, for $\tau \geq 12$ and under condition \fr{nu-value}, as $\varepsilon \rightarrow 0$, one has 
\beqn  \label{r21r31}
 R_{11} + R_{21}=O \left(  \varepsilon^{2}\right)=  O \left( A^2 \left[ \chi_{\varepsilon, A}\right]^{\mu}\right).
\eeqn
Now, combining \fr{r22} and \fr{r32}, and using \fr{var} and \fr{Thres}, one obtains
\beqn  \label{r22r32}
 \Delta=R_{12} + R_{22}&=&  O \left(  \sum^{J_1-1}_{j_1=0}\sum^{J_2-1}_{j_2=0}\sum^{M}_{l=1}\sum_{k, k'}\min \left\{ \left| \theta_{l: \bom}  \right|^2,    {\varepsilon^{2}}\log(1/\varepsilon)l^{-1}\| (\bG^{(l)})^{-1} \|^2\right\} \right)\nonumber\\
 &=&O \left(  \sum^{J_1-1}_{j_1=0}\sum^{J_2-1}_{j_2=0}\sum^{M}_{l=1}\min \left\{ \sum_{k, k'}\left| \theta_{l: \bom}  \right|^2, 2^{j_1 + j_2}   {\varepsilon^{2}}\log(1/\varepsilon)l^{2r-1}\right\} \right).
\eeqn
Then, $\Delta$ can be decomposed into three components, $\Delta_1$, $\Delta_2$ and $\Delta_3$, as follows
\beqn  
 \Delta_1&=&  O \left( \left(\sum^{J_1-1}_{j_1=j_{10}+1} \sum^{J_2-1}_{j_2=0} \sum^{M}_{l=1}+ \sum^{J_1-1}_{j_1=0} \sum^{J_2-1}_{j_2=j_{20} +1} \sum^{M}_{l=1} + \sum^{J_1-1}_{j_1=0} \sum^{J_2-1}_{j_2=0} \sum^{M}_{l=M_0}\right)\sum_{k, k'}\left| \theta_{l: \bom}  \right|^2 \right), \label{del1}\\
 \Delta_2&=&O \left( \sum^{j_{10}}_{j_1=0}  \sum^{j_{20}}_{j_2=0} \sum^{M_0}_{l=1}  A^22^{j_1+ j_2} \left[ \chi_{\varepsilon, A}\right] l^{2r-1 } \II \left(\eta^c_{l: j_1, j_2} \right) \right), \label{del2}\\
 \Delta_3&=& O \left(  \sum^{j_{10}}_{j_1=0}  \sum^{j_{20}}_{j_2=0} \sum^{M_0}_{l=1} \left[\sum_{k, k'}\left| \theta_{l: \bom}  \right|^2\right]  \II \left(\eta_{l: j_1, j_2} \right)\right),\label{del3}
\eeqn
where $\eta_{l: j_1, j_2}= \left\{l, j_1, j_2: 2^{j_1+j_2} l^{2r } >   \left[ \chi_{\varepsilon, A}\right]^{\mu-1} \right\}$. For $\Delta_1$, it is easy to see that for $j_{10}$, $j_{20}$ and $M_0$ given in \fr{Lev:J0} and \fr{Lev:M0}, respectively, 
\beqns
\Delta_1&=&  O \left( A^2 2^{-2j_{10}s_1}+ A^2 2^{-2j_{20}s_2} + A^2 M_0^{-2s_3}\exp \{-2\gamma M_0^{\beta} \}  \right).
\eeqns
Consequently, if $\gamma=\beta=0$, as  $\varepsilon \rightarrow 0$, one has 
\beqn  \label{del111}
 \Delta_1= O \left( A^2 \left[ \chi_{\varepsilon, A}\right]^{\mu}\right).
\eeqn
If $\gamma>0, \beta>0$, then 
\beqn  \label{del11}
 \Delta_1&= &  O \left( A^2 2^{-2j_{10}s_1}+ A^2 2^{-2j_{20}s_2} \right)\nonumber\\
 &=& O \left( A^2 \left[ \chi_{\varepsilon, A}\right]^{\min \left \{ \frac{2s_2}{2s_2+1}, \frac{2s_1}{2s_1+1}\right \} } \right).
\eeqn
For $\Delta_2$ in \fr{del2}, as  $\varepsilon \rightarrow 0$, one obtains
\beqn  \label{d2}
 \Delta_2= O \left( A^2 \left[ A^{-2} {\varepsilon^{2}}\log(1/\varepsilon)\right]\left[ \chi_{\varepsilon, A}\right]^{\mu-1} \right)=O \left( A^2 \left[ \chi^{\alpha}_{\varepsilon, A}\right]^{\mu} \right).
\eeqn
In order to evaluate \fr{del3}, we need to consider five different cases. \\
{\bf Case 1: $\gamma=\beta=0$, $s_1 \leq \min \{s_2, \frac{s_3}{2r} \}$.  }In this case, $\mu = \frac{2s_1}{2s_1 +1}$, \fr{del3} becomes, as   $\varepsilon \rightarrow 0$
\beqn  
 \Delta_3&= &O \left( A^2\sum^{j_{10}}_{j_1=0} \sum^{M_0}_{l=1}\sum^{j_{20}}_{j_2=0}2^{-2 j_1s_1-2 j_2s_2} l^{-2s_3}\II \left(2^{j_1} >   2^{-j_2}\frac{\left[ \chi_{\varepsilon, A}\right]^{\mu -1}}{l^{2r}} \right)\right)\nonumber\\
 &=& O \left( A^2\left[ \chi_{\varepsilon, A}\right]^{ 2 s_1(1-\mu)} \sum^{M_0}_{l=1} l^{-2(s_3-s_12r)}\sum^{j_{20}}_{j_2=0}2^{-2 j_2(s_2-s_1)} \right)\nonumber\\
 &=&O \left( A^2 \left[ \chi_{\varepsilon, A}\right]^{\frac{2s_1}{2s_1 +1}}\left[ \log(\varepsilon^{-1}) \right]^{\II ( s_1=s_2)+ \II ( s_1= s_3/2r)} \right).
 \label{d31}
\eeqn
{\bf Case 2: $\gamma=\beta=0$, $s_2 \leq \min \{s_1, \frac{s_3}{2r} \}$.  }In this case, $\mu = \frac{2s_2}{2s_2 +1}$, \fr{del3} becomes, as   $\varepsilon \rightarrow 0$
\beqn  
 \Delta_3&= &O \left( A^2\sum^{j_{10}}_{j_1=0} \sum^{M_0}_{l=1}\sum^{j_{20}}_{j_2=0}2^{-2 j_1s_1-2 j_2s_2} l^{-2s_3}\II \left(2^{j_2} >   2^{-j_1}\frac{\left[ \chi_{\varepsilon, A}\right]^{\mu -1}}{l^{2r}} \right)\right)\nonumber\\
 &=& O \left( A^2\left[ \chi_{\varepsilon, A}\right]^{ 2 s_2(1-\mu)} \sum^{M_0}_{l=1} l^{-2(s_3-s_22r)}\sum^{j_{10}}_{j_1=0}2^{-2 j_1(s_1-s_2)} \right)\nonumber\\
 &=&O \left( A^2 \left[ \chi_{\varepsilon, A}\right]^{\frac{2s_2}{2s_2 +1}} \left[ \log(\varepsilon^{-1}) \right]^{\II ( s_1=s_2)+ \II ( s_2= s_3/2r)} \right). \label{d32}
\eeqn
{\bf Case 3: $\gamma=\beta=0$, $s_3 \leq \min \{2rs_1, {2rs_2} \}$.  }In this case, $\mu = \frac{2s_3}{2s_3 +2r}$, \fr{del3} becomes, as   $\varepsilon \rightarrow 0$
\beqn  
 \Delta_3&= &O \left( A^2\sum^{j_{10}}_{j_1=0} \sum^{M_0}_{l=1}\sum^{j_{20}}_{j_2=0}2^{-2 j_1s_1-2 j_2s_2} l^{-2s_3}\II \left(l^{2r} >   2^{-j_1-j_2}{\left[ \chi_{\varepsilon, A}\right]^{\mu -1}}{} \right)\right)\nonumber\\
 &=& O \left( A^2\left[ \chi_{\varepsilon, A}\right]^{ -\frac{\mu -1}{2r}2s_3} \sum^{j_{10}}_{j_1=0}2^{-\frac{2 j_1}{2r}(2r s_1-s_3)}\sum^{j_{20}}_{j_2=0}2^{-\frac{2j_2}{2r}(2r s_2 -s_3)} \right)\nonumber\\
   &=&O \left( A^2 \left[ \chi_{\varepsilon, A}\right]^{\frac{s_3}{s_3 + r}} \left[ \log(\varepsilon^{-1}) \right]^{\II ( s_2= s_3/2r)+ \II ( s_1= s_3/2r)}  \right). \label{d33}
\eeqn
{\bf Case 4: $\gamma>0, \beta > 0$, $s_1 \leq s_2$.  }In this case, $\mu = \frac{2s_1}{2s_1 +1}$, \fr{del3} becomes, as   $\varepsilon \rightarrow 0$
\beqn  
 \Delta_3&= &O \left( A^2\sum^{j_{10}}_{j_1=0} \sum^{M_0}_{l=1}\sum^{j_{20}}_{j_2=0}2^{-2 j_1s_1-2 j_2s_2} l^{-2s_3}\exp \{-2\gamma l^{\beta} \}\II \left(2^{j_1} >   2^{-j_2}\frac{\left[ \chi_{\varepsilon, A}\right]^{\mu -1}}{l^{2r}} \right)\right)\nonumber\\
 &=& O \left( A^2\left[ \chi_{\varepsilon, A}\right]^{ \frac{2s_1}{2s_1 +1}} \sum^{j_{20}}_{j_2=0}2^{-2 j_2(s_2-s_1)} \right)\nonumber\\
 &=&O \left( A^2 \left[ \chi_{\varepsilon, A}\right]^{\frac{2s_1}{2s_1 +1}}\left[ \log(\varepsilon^{-1}) \right]^{\II ( s_1=s_2)} \right).\label{delc4}
\eeqn
{\bf Case 5: $\gamma>0, \beta > 0$, $s_2 \leq s_1$.  }In this case, $\mu = \frac{2s_2}{2s_2 +1}$, \fr{del3} becomes, as   $\varepsilon \rightarrow 0$
\beqn  
 \Delta_3&= &O \left( A^2\sum^{j_{10}}_{j_1=0} \sum^{M_0}_{l=1}\sum^{j_{20}}_{j_2=0}2^{-2 j_1s_1-2 j_2s_2} l^{-2s_3}\exp \{-2\gamma l^{\beta} \}\II \left(2^{j_2} >   2^{-j_1}\frac{\left[ \chi_{\varepsilon, A}\right]^{\mu -1}}{l^{2r}} \right)\right)\nonumber\\
 &=& O \left( A^2\left[ \chi_{\varepsilon, A}\right]^{ \frac{2s_2}{2s_2 +1}} \sum^{j_{10}}_{j_1=0}2^{-2 j_1(s_1-s_2)} \right)\nonumber\\
 &=&O \left( A^2 \left[ \chi_{\varepsilon, A}\right]^{\frac{2s_2}{2s_2 +1}}\left[ \log(\varepsilon^{-1}) \right]^{\II ( s_1=s_2)} \right).\label{delc5}
\eeqn
Now, to complete the proof, combine formulae \fr{R'_3}-\fr{delc5}.\\


\section{Introduction to the theory of banded Toeplitz matrices.}
\label{sec:Toeplitz}

The proof of asymptotic optimality of the estimator ${\hat{f}}$
relies heavily on the theory of banded Toeplitz matrices developed in
B\"{o}ttcher  and Grudsky  (2000, 2005). In this subsection, we review some of the
facts about Toeplitz matrices which {were used in the proofs in Section 7}.

Consider a sequence of numbers $\{ b_k \}_{k=-\infty}^\infty$  such that $\sum_{k=-\infty}^\infty |b_k| < \infty$.
An infinite Toeplitz matrix $T=T(b)$ is the matrix with elements $T_{i,j} = b_{i-j}$, $i,j=0,1, \ldots $.

Let ${\cal C} = \{z \in C: |z|=1 \}$ be the complex unit circle.
With each Toeplitz matrix $T(b)$ we can associate its symbol
\be \label{assocfun}
b(z) = \sum_{k=-\infty}^\infty b_k z^k, \ \ z \in {\cal C}.
\ee
Since, $\displaystyle{B(\theta)= b(e^{i\theta}) = \sum_{k=-\infty}^\infty b_k e^{i k \theta}}$,
numbers $b_k$ are Fourier coefficients of function $B(\theta)= b(e^{i\theta})$.
For any function $b(z)$ with an argument on a unit circle ${\cal C}$ denote
$$
\|b\|_{circ} = \displaystyle{\max_{|z|=1} b(z)}.
$$

There is a very strong link between properties of a Toeplitz matrix $T(b)$
and function $b(z)$. In particular, if  $b(z) \neq 0$ for $z \in {\cal C}$
and $\wind (b) = J_b$, then $b(z)$ allows Wiener-Hopf factorization
$b(z) = b_{-} (z)\,  b_{+} (z)\,  z^{J_b}$  where   $b_+$ and $b_-$ have the following forms
$$
b_{-} (z)  =  \sum_{k=0}^\infty b^{-}_{-k} z^{-k}, \ \
b_{+} (z)  =  \sum_{k=0}^\infty b^{+}_{k} z^{k}
$$
(see Theorem 1.8 of B\"{o}ttcher  and Grudsky  (2005)).

If $T(b)$ is a lower triangular Toeplitz matrix, then
$b(z) \equiv b_{+} (z)$ with $b^+_k=b_k$.
In this case,  the product of two Toeplitz matrices can be obtained by simply multiplying their symbols and
the inverse of a Toeplitz matrix can be obtained  by taking the  reciprocal
of function  $b_{+} (z)$:
\be \label{identity}
T(b_{+} d_{+}) = T(b_{+}) T(d_{+}),\ \ \
T^{-1}(b_{+}) = T(1/b_{+}).
\ee

Let $T_m (b) = T_m (b_{+}) \in R^{m \times m}$ be a banded lower triangular
Toeplitz matrix corresponding to the Laurent polynomial
$\displaystyle{b  (z) = \sum_{k=0}^{m-1} b_k z^k}$.

In practice, one usually use only finite, banded, Toeplitz matrices with
elements $T_{i,j}$, $i,j=0,1, \ldots, m-1$.  In this case,  only a finite  number
of coefficients $b_k$ do not vanish and function $b(z)$ in \fr{assocfun}
reduces to a  Laurent polynomial $\displaystyle{b(z) = \sum_{k=-J}^{K} b_k z^k}$,
 $z \in {\cal C}$,  where $J$ and $K$ are nonnegative integers, $b_{-J} \neq 0$ and  $b_{K} \neq 0$.
If $b(z) \neq 0$ for $z \in {\cal C}$, then $b(z)$ can be represented in a form
\be \label{Laurent_pol}
b(z) = z^{-J} b_{K} \prod_{j=1}^{J_0} (z-\mu_j) \prod_{k=1}^{K_0} (z - \nu_k) \ \ \mbox{with}\ \
|\mu_j|<1,\, |\nu_k|>1.
\ee
In this case, the winding number of $b(z)$ is $\wind (b) = J_0 - J$.

Let $T_m (b) = T_m (b_{+}) \in R^{m \times m}$ be a banded lower triangular
Toeplitz matrix corresponding to the Laurent polynomial
$\displaystyle{b  (z) = \sum_{k=0}^{m-1} b_k z^k}$.
If $b$ has no zeros on the complex unit circle ${\cal C}$   and $\wind(b) =0$, then, due to Theorem 3.7
 of  B\"{o}ttcher  and Grudsky  (2005),  $T(b)$ is invertible and
$\displaystyle{\lim_{m \rightarrow \infty} \sup \rho(T_m^{-1} (b)) < \infty}$.
Moreover, by Corollary 3.8,
\be \label{norm_converg}
\lim_{m \rightarrow \infty} \rho(T_m^{-1} (b)) = \rho(T^{-1} (b))
\ee

In the paper, we need the following result that is a combination of Lemmas 3 and 4 of Comte {\it et al.} (2017).

\begin{lemma} \label{lem:comte_lem3_4} 
Let function $g$  in \fr{eq1} satisfy Assumptions {\bf A1} and {\bf A2}. Then, 
$G^{(L)} =  T_L((1-z)^r v(z))$ where function $v(z)$ has all its zeros outside the complex unit circle,
so that  $\|T_L(v(z))\|^2 = \lam_{\max} [T_L^T(v(z)) T_L(v(z))] < \|v\|_{circ}^2 < \infty$.
\end{lemma}


\end{document}